\documentclass[prl,twocolumn,superscriptaddress,showpacs,preprintnumbers,onlyamsmath,amssymb]{revtex4}
\usepackage{graphicx}
\usepackage{dcolumn}
\usepackage{bm}

\newcommand{\cyclo}{$ \omega_{\rm c} $}
\newcommand{\ns}{$ N_{\rm s} $}

\newcommand{\gaas}{$\rm GaAs-Al_{x}Ga_{1-x}As$}

\begin{document}

 \title{Density dependence of microwave induced magneto-resistance oscillations in a two-dimensional electron gas}

 \author{B.Simovi\v c}
 \affiliation{Laboratorium f\"{u}r Festk\"{o}rperphysik, Eidgen\"{o}ssische
Technische Hochschule, CH-8093 Z\"{u}rich}
 \author{C. Ellenberger}
  \affiliation{Laboratorium f\"{u}r Festk\"{o}rperphysik, Eidgen\"{o}ssische
Technische Hochschule, CH-8093 Z\"{u}rich}
\author{K.Ensslin}
  \affiliation{Laboratorium f\"{u}r Festk\"{o}rperphysik, Eidgen\"{o}ssische
Technische Hochschule, CH-8093 Z\"{u}rich}
\author{W.Wegscheider}
  \affiliation{Institute f\"{u}r Experimentelle und Angewandte Physik, Universit\"{a}t Regenburg, D-93040 Regenburg, Germany}


 \begin{abstract}
We have measured the magneto-resistance of a two-dimensional
electron gas (2DEG) under continuous microwave irradiation as a
function of electron density and mobility tuned with a metallic
top-gate. In the entire range of density and mobility we have
investigated, we observe  microwave induced oscillations of large
amplitude that are B-periodic. These B-periodic oscillations are
reminiscent of the ones reported by Kukushkin \textit{et al
}~\cite{Kuku:NewTy} and  which were attributed to the presence of
edge-magneto-plasmons. We have found that the B-periodicity does
not increase linearly with the density in our sample  but shows a
plateau in the range (2.4-3)$\times 10^{11}\rm cm^{-2} $. In this
regime, the phase of the B-periodic oscillations is found to shift
continuously by two periods.
\end{abstract}

 \pacs{73.43.Qt, 73.40.-c, 75.47.-m}

\maketitle

The recent discovery of the vanishing of electrical resistance in
a ultra-high mobility 2DEG under microwave irradiation has sparked
a large interest in the photoresponse of quantum Hall systems. The
so-called microwave-induced "zero-resistance" states have been
observed in the magnetic field regime where the cyclotron and
microwave frequencies, \cyclo\ and $\omega$ respectively, are such
that \cyclo$\leq \omega$.~\cite{Zudov:EvidNew,Mani:ZeroR}. A
key-aspect of this phenomenon lies in its strong dependence on the
mobility. In particular, for $\mu = 3 \times 10 ^{6} \rm cm^{2}/ V
s $, some oscillations with a periodicity determined by the ratio
$\omega /\omega_{\rm c}$ are discerned  in  the
magneto-transport~\cite{Zudov:ShubdH} while for $\mu = 2.5\times
10 ^{7} \rm cm^{2}/ V s $, these minima fully drop to zero,
thereby revealing a sharp suppression of the dissipative processes
in the 2DEG.~\cite{Zudov:EvidNew,Mani:ZeroR}

Independently of this discovery, a different type of microwave
related effects, also leading to a strong modulation of the
magneto-resistance, has been recently reported to occur in samples
with a moderately high mobility of $\mu = 1.3 \times 10 ^{6} \rm
cm^{2}/ V s $.~\cite{Kuku:NewTy}  In contrast to the 1/B-periodic
oscillations mentioned above, these are B-periodic and develop in
the field range where $\omega \leq \omega_{\rm c}$. The same
experiment performed on samples of different sizes and electron
densities has shown that the period of these B-periodic
oscillations increases with density and is inversely proportional
to the length of the Hall bar.  As discussed in
ref~\cite{Kuku:NewTy}, these features can be reasonably understood
on the basis of edge-magneto-plasmon excitations.  Several
scenarios for the understanding of the 1/B-periodic oscillations
have been put forward
recently.~\cite{And:dyn,Durst:radI,Jun:rad,Kou:Cla,Lei:Rad,Ryzh:Mec,Ryzhhii:MicroP,Ryzhii:NonLinear,Dmit:OscAc,Dmitr:Cyclo}
It is unclear whether the two effects, namely the B-periodic and
the 1/B-periodic oscillations, can coexist in one sample, for
there has been no report so far on their coexisting together and
to the best of our knowledge, the issue has not yet been
addressed. Some recent theoretical and experimental
works~\cite{Mikha:MicroInd, stud:micoAbs} have stressed the
importance of considering  plasmon-related excitations to account
for the properties of the 1/B-oscillations but a clear
understanding of the electrodynamics at work  in these systems is
still lacking.

In this report, we present a study of the  microwave-induced
oscillations versus electron density and mobility in a  Hall bar
patterned on a high-mobility \gaas\ heterostructure. In contrast
to previous experiments where the values of \ns\ and $\mu$ were
obtained from a brief exposure to light at low temperature, \ns\
and $\mu$ here are tuned continuously by adjusting a top-gate
voltage. This permits a systematic study of microwave-induced
effects in a large range of densities and mobilities.  By
increasing the electron density from $0.5$ to $5\times 10^{11} \rm
cm^{-2}$, we could tune the mobility in the range of $(1-7) \times
10^{6} \rm cm^{2}/ V s$. In the entire range of densities and
mobilities investigated we have observed oscillations of large
amplitude that are B-periodic accompanied with a dependence to
light exposure. Furthermore, the period of the oscillations does
not increase linearly with \ns\ but shows unexpectedly a plateau
in the range of (2.4-3)$\times 10^{11}\rm cm^{-2} $ which occurs
concomitantly with a continuous phase shifting of the oscillations
by two periods.

\begin{figure}[t!]
 \vspace{0cm}
\begin{center}
\includegraphics[width=0.8\linewidth]{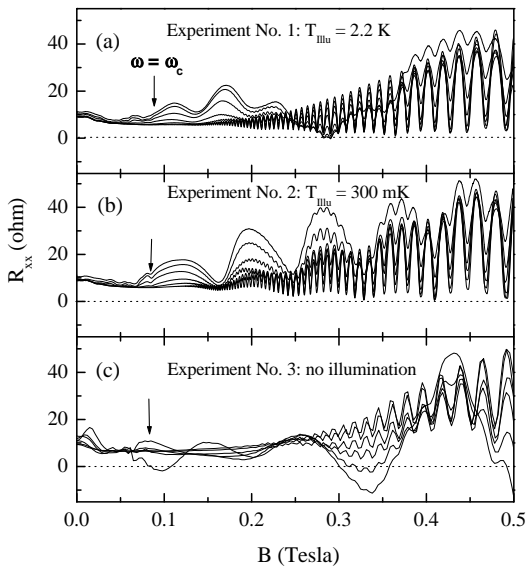}
\caption{ Magneto-resistance measured under continuous microwave
irradiation for different powers is compared for three different
illumination conditions. The microwave power P lies in the range
(1$\mu \rm W$-1 mW). \ns=$5\times 10^{11} \rm cm^{-2}$ and $\mu
=7.50 \times 10^{6} \rm cm^{2}/ V \ s $   for panel(a) and (b)
while \ns=$4\times 10^{11} \rm cm^{-2}$ and $\mu =5.3\times 10^{6}
\rm cm^{2}/ V \ s $ for panel (c)}\label{all}
\end{center}
\end{figure}

\begin{figure}[tb]
 \vspace{0cm}
\begin{center}
\includegraphics[width=0.8\linewidth]{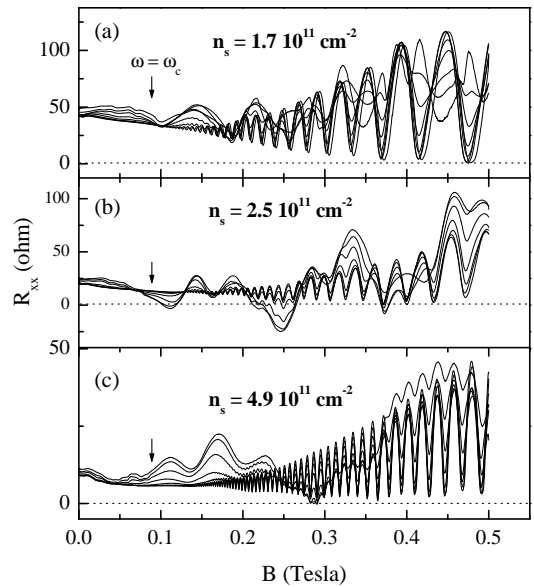}
\caption{Experiment No.~1: illumination at $T_{\rm Illu} = 2.2
\rm K$. Magneto-resistance measured for three different densities
versus the microwave power P in the range (1$\mu \rm W$-1 mW). The
value of the B-field at which \cyclo=$\omega$ is marked with an
arrow. }\label{expSDH}
\end{center}
\end{figure}
Our sample is a high-mobility ($\mu \geqslant 1.4 \times 10^{7}
\rm cm ^{2}/ Vs)$ two-dimensional electron gas in \gaas\
heterostructures grown by molecular-beam epitaxy. The 2DEG is
located 100nm below the sample surface. This high mobility is
obtained after  brief illumination with a red light-emitting diode
on a macroscopic Van der Pauw geometry. After the Hall bar has
been lithographically defined, the mobility drops significantly,
its maximal value depending on experimental conditions. The Hall
bar is $200\mu m$ long and $50 \mu m$ wide. A TiAu top-gate (100
nm) has been evaporated on the Hall bar structure in order to tune
the density and the mobility independently of illumination.
Increasing the top-gate voltage from -100 mV to 640 mV increases
the electron density $N_{\rm s}$ and the mobility $\mu$ from 0.5
to $5\times 10^{11} \rm cm^{-2}$ and 0.5 to $7.50 \times 10^{6}
\rm cm^{2}/ V \ s$ respectively. \ns\ was found to vary linearly
with the top-gate voltage within this range. The mobility
increases linearly up to \ns$=3\times 10^{11} \rm cm^{-2}$ and
gradually reaches saturation above. A signal generator operating
at 0.001-40 GHz, with typical output power from 10 to 0.1 mW  was
used as the source of microwaves. They were guided into a dilution
fridge via an oversized waveguide with an attenuation of 5dB and
the sample was placed in the near field of the waveguide aperture.
The frequency was fixed to 35.5GHz for all experiments. The values
of microwave power mentioned in what follows are the power levels
estimated at the sample location. All data presented here are
obtained in the regime where the measured voltages are linear in
the applied sinusoidal current (21 Hz) ranging between 50 and 200
nA. The magneto-resistance $R_{\rm xx}$ was measured under
continuous microwave irradiation while the photovoltage $V_{\rm
xx}$, which develops across the Hall bar independently of the
applied current was measured via dc and ac techniques using for
the latter a 2kHz-ac-modulation of the microwave power. The two
ways of measuring provided consistent results.

\begin{figure}[tb]
 \vspace{0cm}
\begin{center}
\includegraphics[width=0.7\linewidth]{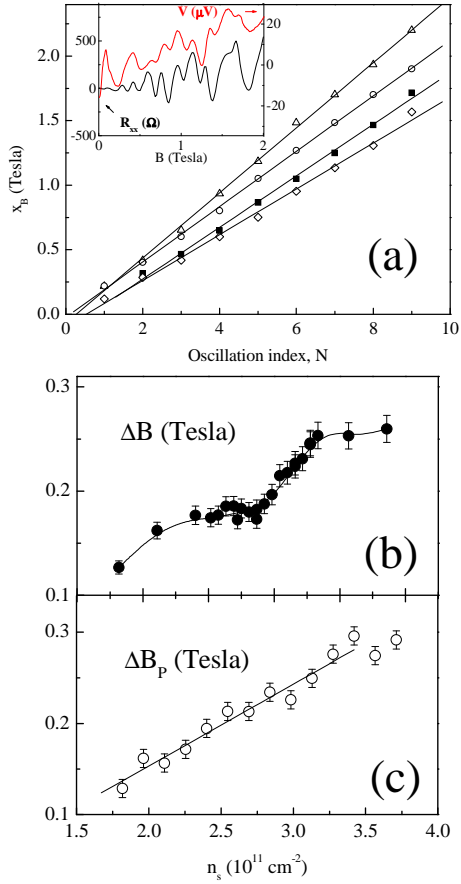}
\caption{Experiment No.~3. Panel(a): position of the maxima vs.
oscillation index for various density ranging between 2.4 and
$3.6\times 10^{11} \rm cm^{-2}$. The period $\Delta B$ is deduced
from a linear fit  with a reliability factor fluctuating between
99.7 and 99.9 \%. Inset: magneto-resistance (continuous line) and
photovoltage (dotted line) measured for \ns = $2.4\times 10^{11}
\rm cm^{-2}$. The current through the Hall bar is 200nA and P $
\sim 300\mu \rm W$. Panel (b): period of the photoresistance vs.
\ns. Panel (c): period of the photovoltage vs.
\ns.}\label{exp3density}
\end{center}
\end{figure}

\begin{figure}[t!]
 \vspace{0cm}
\begin{center}
\includegraphics[width=0.6\linewidth]{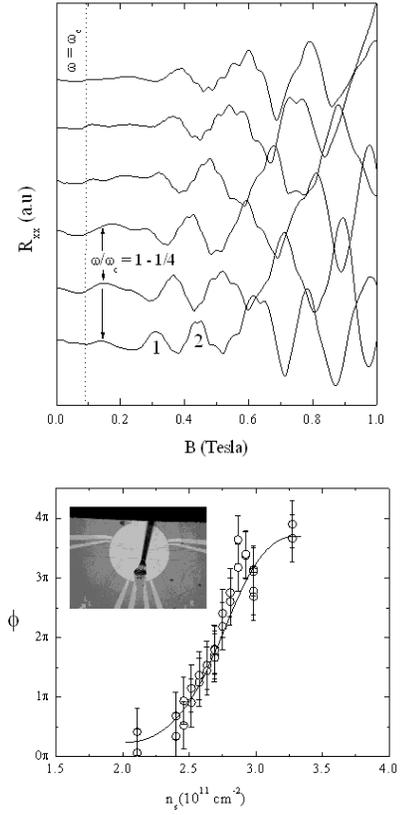}
\caption{Experiment No.~3. Panel (a): microwave-induced
oscillations in the low field range measured for various \ns\
ranging between 2.4 and $2.9\times 10^{11} \rm cm^{-2}$. The
curves are offset for clarity and correspond to a step upward in
density of $ 0.1\times 10^{11} \rm cm^{-2}$ from the lowest one.
Panel(b) show the phase $\phi$ of the B-periodic oscillations
versus \ns. The phase was determined from the denoted 1 and 2
maxima in panel(a) in respect to the B-field at which
\cyclo$=\omega$ i.e 0.09T for $\omega = 35.5 \rm GHz$. The
continuous line is a guide for the eyes. Inset: picture of the
Hall-bar sample with its TiAu top-gate (circle). The distance
between the voltage probes is 200 $\mu m$}\label{exp3all}
\end{center}
\end{figure}

The data  can be divided into three distinct experiments which
differ in the condition of exposure to a red light emitting diode.
For the sake of clarity we refer to these three experiments as
experiment No.~1, 2 and 3 respectively. For experiment No.~1, the
sample was briefly illuminated at the temperature T =2.2K and then
cooled down to 100 mK which is the base temperature we could reach
with our set-up. Note that with microwaves on, the mixing chamber
of our dilution fridge heats up to T = 300mK for the maximum power
investigated. Experiment No.2 was performed during the same cool
down. The 2DEG was first depleted by applying a negative top-gate
voltage and then charged  again and exposed to a brief red light
illumination at T = 300 mK. Experiment No.~3, however, involves a
different cool down during  which the sample was not exposed to
illumination at all. In Fig.~1 we  compare  the longitudinal
magneto-resistance observed for the three experiments at
comparable \ns\ for various  microwave powers ranging between 1
$\mu \rm W$ and 1 mW. In each case, oscillations of large
amplitude induced by microwave irradiation develop on top of
Shubnikov de Haas (SDH) oscillations starting at about the
threshold where \cyclo$> \omega$. It is obvious, however, that the
properties of the B-periodic oscillations are strongly affected by
the exposure to light. Their period $\Delta B$ and their damping
with magnetic field differ strongly while \ns\ is the same in
experiment No.~1 and 2 and only 20\% less for experiment No.~3. In
addition, we see in the panel (c) of Fig.~1 that the sign of
$\Delta R_{\rm xx}$ can be negative. This is not specific to
experiment No.~3 but was also observed in experiment No.~1 by
lowering the electron density as shown in Fig.~2. The B-periodic
oscillations can therefore lead to an increased, a reduced or a
negative absolute longitudinal resistance depending on the details
of the parameter setting. Also, when scaling the amplitude of the
oscillations in Fig.~2 to the value of $R_{\rm xx}$ without
microwaves, we have observed (not shown) that the magnitude of the
effect increases with electron mobility and saturates for
microwave powers beyond 500 $\mu \rm W$.

We have seen that the  detailed features of the oscillations
depend significantly on how the sample has been exposed to red
light. Independently of these variations, however, their behavior
is reminiscent of the B-periodic oscillations observed in
ref~\cite{Kuku:NewTy} and which were attributed to
edge-magneto-plasmon excitations. Also like in
ref~\cite{Kuku:NewTy}, we have observed that a photovoltage
develops across the Hall bar which shows a  similar periodicity. A
striking fact is that there is no clear feature of the
1/B-periodic oscillations for \cyclo$<\omega$ in Fig.~1 and 2
despite the high mobilities achieved.

 A more detailed study of the effect versus density has been
carried out and the results are summarized in Fig.3 and 4. We
first show the position $x_{\rm B}$ of the maxima of the
oscillations as a function of their oscillation index N in
Fig.~3a. The B-periodicity is clear from the linear dependence and
as seen in the inset, the oscillations can be followed up to 2
Tesla in the quantum Hall regime, the resistance oscillations
being linear in the applied current. Under the present
experimental conditions where the resistance is measured with ac
current modulation the resistance signal is not modified by the dc
photovoltage excited by the unmodulated microwave.
\newline The period $\Delta B$ of the
resistance as well as the period $\Delta B_{\rm P}$ of the
photovoltage are shown vs. \ns\ in Fig.~3b and 3c respectively. In
ref~\cite{Kuku:NewTy}, the two were found to be equal and to
increase linearly with \ns. In the present work, however, it can
be seen in Fig.~3 that they behave differently upon varying \ns.
On one hand, $\Delta B_{\rm P}$ increases linearly with \ns, on
the other hand,  $\Delta B$ shows a step increase through a narrow
plateau for \ns\ between 2.4 and $2.9\times 10^{11} \rm cm^{-2}$.

The details of the oscillations at low field are displayed in
Fig.~4a for very closely-spaced values of top-gate voltage.  We
see that the phase shifts considerably in a narrow voltage-range
corresponding only to a 4\% variation in \ns. This phase shifting,
also shown in Fig.~4b, does not occur in the photovoltage signal
and is accompanied with the emergence of a broad maximum located
at $B\approx 0.11 \ \rm Tesla$ which belongs to a separate pattern
than the B-periodic oscillations. As seen in Fig.~4a, this
additional peak  can be resolved from the B-periodic oscillations
for \ns\ ranging between 2.4 and $2.9\times 10^{11} \rm cm^{-2}$,
thereby indicating the sharpness of the feature in term of
electron density. Its location would correspond to the first
cyclotron resonance harmonic at $\omega/\omega_{c} = 1-1/4$
expected for the 1/B-periodic oscillations. This may suggest that
the two phenomena, namely B-periodic and 1/B-periodic
oscillations, could coexist in a very narrow range of density and
mobility. We expect indeed to observe these 1/B-periodic
oscillations, for  $\mu \geq 4 \times 10^{6} \rm cm^{2}/ V s$ in
Fig.~4a which is a value comparable to $\mu \geq 3 \times 10^{6}
\rm cm^{2}/ V s$ as reported in ref~\cite{Zudov:ShubdH}. Yet, the
presence of only one multiplicity would mean an anomalously large
damping of the 1/B-periodic oscillations in our sample which
remains to be explained.

There have been so far no report concerning the phase of the
B-periodic oscillations. Nevertheless, the fact that the plateau
in $\Delta B$ and the phase shifting occur together, lends support
to the idea that these two features are connected. This hints at
the presence of hindered microwave absorption mechanisms in our
system and some questions arise as to what extent they could be
related to the microscopic mechanisms underlying the 1/B-periodic
oscillations. The emergence of a broad maximum located where
$\omega/\omega_{c} = 1-1/4$ in the plateau regime is indeed
suggestive of a possible cross-talk between the
edge-magneto-plasmons and the phenomenon of 1/B-periodic
oscillations but the lack of any clear pattern of maxima in the
range \cyclo$< \omega$ limits the discussion to this point.

An important result of our experiments is that the oscillations simultaneously observed in
 the resistance and the photovoltage, while both being B-periodic, may differ in phase and
 period in certain density regimes. The inset of Fig. 4(b) shows an optical microscope image
 of the gate covering the Hall bar. It is conceivable that magnetoplasmons are excited
 differently in the Hall bar (resistance signal) and in the contact regions outside of the
  Hall bar not covered by the gate (photovoltage).

To conclude, we have presented a study of the microwave-induced
oscillations in 2DEG as a function of electron density and
mobility. We have observed a \textsl{clear} pattern of B-periodic
oscillations for \cyclo$>\omega$, the properties of which  such as
amplitude and period vary significantly for different exposure
conditions to red light. A careful study of these oscillations
versus electron density  shows that the period does not increase
linearly with density as it was initially expected from a
classical picture based on edge-magneto-plasmon excitations.
Rather, the period shows a plateau in the range (2.4-2.9)$\times
10^{11}\rm cm^{-2} $ together with a continuous phase shifting by
two periods. These sharp features could be related to the
coexistence of strongly damped 1/B-periodic oscillations and
B-periodic oscillations in a narrow range of density and mobility
although no quantitative explanation for such effect can be
derived at this stage.

\bibliographystyle{apsrev}
\bibliography{2DEG_microwave}

\begin{thebibliography}{16}
\expandafter\ifx\csname natexlab\endcsname\relax\def\natexlab#1{#1}\fi
\expandafter\ifx\csname bibnamefont\endcsname\relax
  \def\bibnamefont#1{#1}\fi
\expandafter\ifx\csname bibfnamefont\endcsname\relax
  \def\bibfnamefont#1{#1}\fi
\expandafter\ifx\csname citenamefont\endcsname\relax
  \def\citenamefont#1{#1}\fi
\expandafter\ifx\csname url\endcsname\relax
  \def\url#1{\texttt{#1}}\fi
\expandafter\ifx\csname urlprefix\endcsname\relax\def\urlprefix{URL }\fi
\providecommand{\bibinfo}[2]{#2}
\providecommand{\eprint}[2][]{\url{#2}}

\bibitem[{\citenamefont{Kukushkin et~al.}(2004)\citenamefont{Kukushkin, Akimov,
  Smet, Mikhailov, von Klitzing, Aleiner, and Falko}}]{Kuku:NewTy}
\bibinfo{author}{\bibfnamefont{I.-V.} \bibnamefont{Kukushkin}},
  \bibinfo{author}{\bibfnamefont{M.-Y.} \bibnamefont{Akimov}},
  \bibinfo{author}{\bibfnamefont{J.-H.} \bibnamefont{Smet}},
  \bibinfo{author}{\bibfnamefont{S.-A.} \bibnamefont{Mikhailov}},
  \bibinfo{author}{\bibfnamefont{K.}~\bibnamefont{von Klitzing}},
  \bibinfo{author}{\bibfnamefont{I.-L.} \bibnamefont{Aleiner}},
  \bibnamefont{and} \bibinfo{author}{\bibfnamefont{V.-I.} \bibnamefont{Falko}},
  \bibinfo{journal}{Phys. Rev. Lett.} \textbf{\bibinfo{volume}{92}},
  \bibinfo{pages}{236803} (\bibinfo{year}{2004}).

\bibitem[{\citenamefont{Zudov et~al.}(2003)\citenamefont{Zudov, Du, Pfeiffer,
  and West}}]{Zudov:EvidNew}
\bibinfo{author}{\bibfnamefont{M.-A.} \bibnamefont{Zudov}},
  \bibinfo{author}{\bibfnamefont{R.-R.} \bibnamefont{Du}},
  \bibinfo{author}{\bibfnamefont{L.-N.} \bibnamefont{Pfeiffer}},
  \bibnamefont{and} \bibinfo{author}{\bibfnamefont{K.-W.} \bibnamefont{West}},
  \bibinfo{journal}{Phys. Rev. Lett.} \textbf{\bibinfo{volume}{90}},
  \bibinfo{pages}{046807} (\bibinfo{year}{2003}).

\bibitem[{\citenamefont{Mani et~al.}(2002)\citenamefont{Mani, Smet, von
  Klitzing, Harayanamurti, Johnson, and Umansky}}]{Mani:ZeroR}
\bibinfo{author}{\bibfnamefont{R.-G.} \bibnamefont{Mani}},
  \bibinfo{author}{\bibfnamefont{J.-H.} \bibnamefont{Smet}},
  \bibinfo{author}{\bibfnamefont{K.}~\bibnamefont{von Klitzing}},
  \bibinfo{author}{\bibfnamefont{V.}~\bibnamefont{Harayanamurti}},
  \bibinfo{author}{\bibfnamefont{W.-B.} \bibnamefont{Johnson}},
  \bibnamefont{and} \bibinfo{author}{\bibfnamefont{V.}~\bibnamefont{Umansky}},
  \bibinfo{journal}{Nature} \textbf{\bibinfo{volume}{420}},
  \bibinfo{pages}{646} (\bibinfo{year}{2002}).

\bibitem[{\citenamefont{Zudov et~al.}(2001)\citenamefont{Zudov, Du, Simmons,
  and Reno}}]{Zudov:ShubdH}
\bibinfo{author}{\bibfnamefont{M.-A.} \bibnamefont{Zudov}},
  \bibinfo{author}{\bibfnamefont{R.-R.} \bibnamefont{Du}},
  \bibinfo{author}{\bibfnamefont{J.-A.} \bibnamefont{Simmons}},
  \bibnamefont{and} \bibinfo{author}{\bibfnamefont{J.-L.} \bibnamefont{Reno}},
  \bibinfo{journal}{Phys. Rev. B} \textbf{\bibinfo{volume}{64}},
  \bibinfo{pages}{201311} (\bibinfo{year}{2001}).

\bibitem[{\citenamefont{Andreev et~al.}(2003)\citenamefont{Andreev, Aleiner,
  and Millis}}]{And:dyn}
\bibinfo{author}{\bibfnamefont{A.-V.} \bibnamefont{Andreev}},
  \bibinfo{author}{\bibfnamefont{I.-L.} \bibnamefont{Aleiner}},
  \bibnamefont{and} \bibinfo{author}{\bibfnamefont{A.-J.}
  \bibnamefont{Millis}}, \bibinfo{journal}{Phys. Rev. Lett.}
  \textbf{\bibinfo{volume}{91}}, \bibinfo{pages}{056803}
  (\bibinfo{year}{2003}).

\bibitem[{\citenamefont{Durst et~al.}(2003)\citenamefont{Durst, Sachdev, Read,
  and Girvin}}]{Durst:radI}
\bibinfo{author}{\bibfnamefont{A.-C.} \bibnamefont{Durst}},
  \bibinfo{author}{\bibfnamefont{S.}~\bibnamefont{Sachdev}},
  \bibinfo{author}{\bibfnamefont{N.}~\bibnamefont{Read}}, \bibnamefont{and}
  \bibinfo{author}{\bibfnamefont{S.-M.} \bibnamefont{Girvin}},
  \bibinfo{journal}{Phys. Rev. Lett.} \textbf{\bibinfo{volume}{91}},
  \bibinfo{pages}{086803} (\bibinfo{year}{2003}).

\bibitem[{\citenamefont{Junren-Shi and Xie}(2003)}]{Jun:rad}
\bibinfo{author}{\bibnamefont{Junren-Shi}} \bibnamefont{and}
  \bibinfo{author}{\bibfnamefont{X.-C.} \bibnamefont{Xie}},
  \bibinfo{journal}{Phys. Rev. Lett.} \textbf{\bibinfo{volume}{91}},
  \bibinfo{pages}{086801} (\bibinfo{year}{2003}).

\bibitem[{\citenamefont{Koulakov and Raikh}(2003)}]{Kou:Cla}
\bibinfo{author}{\bibfnamefont{A.-A.} \bibnamefont{Koulakov}} \bibnamefont{and}
  \bibinfo{author}{\bibfnamefont{M.-E.} \bibnamefont{Raikh}},
  \bibinfo{journal}{Phys. Rev. B} \textbf{\bibinfo{volume}{68}},
  \bibinfo{pages}{115324} (\bibinfo{year}{2003}).

\bibitem[{\citenamefont{Lei and Liu}(2003)}]{Lei:Rad}
\bibinfo{author}{\bibfnamefont{X.-L.} \bibnamefont{Lei}} \bibnamefont{and}
  \bibinfo{author}{\bibfnamefont{S.-Y.} \bibnamefont{Liu}},
  \bibinfo{journal}{Phys. Rev. Lett.} \textbf{\bibinfo{volume}{91}},
  \bibinfo{pages}{226805} (\bibinfo{year}{2003}).

\bibitem[{\citenamefont{Ryzhii et~al.}(2004)\citenamefont{Ryzhii, Suris, and
  Shchamkhalova}}]{Ryzh:Mec}
\bibinfo{author}{\bibfnamefont{V.}~\bibnamefont{Ryzhii}},
  \bibinfo{author}{\bibfnamefont{R.}~\bibnamefont{Suris}}, \bibnamefont{and}
  \bibinfo{author}{\bibfnamefont{B.}~\bibnamefont{Shchamkhalova}},
  \bibinfo{journal}{Physica E} \textbf{\bibinfo{volume}{22}},
  \bibinfo{pages}{13} (\bibinfo{year}{2004}).

\bibitem[{\citenamefont{Ryzhii}(2003)}]{Ryzhhii:MicroP}
\bibinfo{author}{\bibfnamefont{V.}~\bibnamefont{Ryzhii}},
  \bibinfo{journal}{Phys. Rev. B} \textbf{\bibinfo{volume}{68}},
  \bibinfo{pages}{193402} (\bibinfo{year}{2003}).

\bibitem[{\citenamefont{Ryzhii and Suris}(2003)}]{Ryzhii:NonLinear}
\bibinfo{author}{\bibfnamefont{V.}~\bibnamefont{Ryzhii}} \bibnamefont{and}
  \bibinfo{author}{\bibfnamefont{R.}~\bibnamefont{Suris}}, \bibinfo{journal}{J.
  Phys.} \textbf{\bibinfo{volume}{15}}, \bibinfo{pages}{6855}
  (\bibinfo{year}{2003}).

\bibitem[{\citenamefont{Dmitriev et~al.}(2004)\citenamefont{Dmitriev, Mirlin,
  and Polyakov}}]{Dmit:OscAc}
\bibinfo{author}{\bibfnamefont{I.-A.} \bibnamefont{Dmitriev}},
  \bibinfo{author}{\bibfnamefont{A.-D.} \bibnamefont{Mirlin}},
  \bibnamefont{and} \bibinfo{author}{\bibfnamefont{D.-G.}
  \bibnamefont{Polyakov}}, \bibinfo{journal}{Phys. Rev. B}
  \textbf{\bibinfo{volume}{70}}, \bibinfo{pages}{165305}
  (\bibinfo{year}{2004}).

\bibitem[{\citenamefont{Dmitriev et~al.}(2003)\citenamefont{Dmitriev, Mirlin,
  and Polyakov}}]{Dmitr:Cyclo}
\bibinfo{author}{\bibfnamefont{I.-A.} \bibnamefont{Dmitriev}},
  \bibinfo{author}{\bibfnamefont{A.-D.} \bibnamefont{Mirlin}},
  \bibnamefont{and} \bibinfo{author}{\bibfnamefont{D.-G.}
  \bibnamefont{Polyakov}}, \bibinfo{journal}{Phys. Rev. Lett.}
  \textbf{\bibinfo{volume}{91}}, \bibinfo{pages}{226802}
  (\bibinfo{year}{2003}).

\bibitem[{\citenamefont{Mikhailov}(2004)}]{Mikha:MicroInd}
\bibinfo{author}{\bibfnamefont{S.-A.} \bibnamefont{Mikhailov}},
  \bibinfo{journal}{Phys. Rev. B} \textbf{\bibinfo{volume}{70}},
  \bibinfo{pages}{165311} (\bibinfo{year}{2004}).

\bibitem[{\citenamefont{Studenikin et~al.}(2004)\citenamefont{Studenikin,
  Potemski, Sachrajda, Hilke, Pfeiffer, and West}}]{stud:micoAbs}
\bibinfo{author}{\bibfnamefont{S.-A.} \bibnamefont{Studenikin}},
  \bibinfo{author}{\bibfnamefont{M.}~\bibnamefont{Potemski}},
  \bibinfo{author}{\bibfnamefont{A.-S.} \bibnamefont{Sachrajda}},
  \bibinfo{author}{\bibfnamefont{M.}~\bibnamefont{Hilke}},
  \bibinfo{author}{\bibfnamefont{L.-N.} \bibnamefont{Pfeiffer}},
  \bibnamefont{and} \bibinfo{author}{\bibfnamefont{K.-W.} \bibnamefont{West}},
  \bibinfo{journal}{cond-mat/0403058}  (\bibinfo{year}{2004}).

\end{thebibliography}

\end{document}